\documentclass{ws-p10x7}

\usepackage{rotating}

\begin{document}

\title{Electron-Positron- Colliders}

\author{R.-D. Heuer}

\address{Institut f\"ur Experimentalphysik, Universit\"at Hamburg,
Luruper Chaussee 149, 22761 Hamburg
Germany\\E-mail: rolf-dieter.heuer@desy.de}

\twocolumn[\maketitle\abstract
{An electron-positron linear collider in the energy range between 500 and
1000 GeV is of crucial importance 
to precisely test the Standard Model and to explore the physics beyond it.
The physics program is complementary to that of the Large
Hadron Collider.  
Some of the main physics goals and the expected accuracies of the anticipated
measurements at such a linear collider are discussed. 
A short review of the different collider designs presently under study is
given including possible upgrade paths to the multi-TeV region. 
Finally a framework is presented
within which the realisation of such a project could be achieved as a global
international project. }]

\section{Introduction}

A coherent picture of matter and forces
has emerged in the past decades through intensive
theoretical and experimental studies. It is adequately described by the 
Standard Model of particle physics. 
In the last few years many aspects of the model have been stringently
tested, some to the per-mille level, with $e^+e^-$, $ep$ and $p\bar p$ 
machines making complementary contributions, especially to the determination
of the electroweak parameters. Combining the results with neutrino scattering
data and low energy measurements, the experimental analysis is in excellent 
concordance with the electroweak part of the Standard Model. Also the
predictions of QCD have been thoroughly tested, examples being precise
measurements of the strong coupling $\alpha_s$ and probing the proton 
structure to the shortest possible distances. \\
Despite these great successes there are many gaps in our understanding.
The clearest one is the present lack of any direct evidence for the
dynamics of electroweak symmetry breaking and the generation of the
masses of gauge bosons and fermions. The Higgs mechanism which generates
the masses of the fundamental particles in the Standard Model, has not
been experimentally established though the indirect evidence from
precision measurements is very strong.
Even if successfully completed, the Standard Model does not provide
a comprehensive theory of matter. There is no explanation for the wide
range of masses of the fermions, 
the grand unification between the two gauge theories,
electroweak and QCD, is not realised and gravity is not incorporated at the
quantum level. \\
Several alternative scenarios have been developed for the physics which
may emerge beyond the Standard Model as energies are increased.
The Supersymmetric extension of the Standard Model provides a stable
bridge from the presently explored energy scales up to the grand
unification scale. Alternatively, new strong interactions give rise to 
strong forces
between $W$ bosons at high energies. Quite general arguments suggest
that such new phenomena must appear below a scale of approximately 3 TeV. 
Extra space dimensions which alter the high energy behaviour in such a way
that the energy scale of gravity is in the same order as the electroweak
scale are another proposed alternative. \\

There are two ways of exploring the new scales, through attaining the
highest possible energy in a hadron collider and through high precision
measurements at the energy frontier of lepton colliders. \\

This article is based on the results of many workshops on physics and
detector studies for linear colliders. Much more can be found in the
respective publications~\cite{ecfaphys,snowmass,jlcphys,fnal}
and on the different Web sites~\cite{world,acfa,ecfa,usweb}.
Many people have contributed to these studies and the references to their
work can be found in the documents quoted above.

\section{Complementarity of Lepton and Hadron Machines}

It is easier to accelerate protons to very high energies
than leptons, but the 
detailed collision process cannot be well controlled or selected.  
Electron-positron colliders offer a well defined initial state. The
collision energy $\sqrt{s}$ is known and it is tuneable thereby
allowing the choice of the best suited centre-of-mass energy, e.g.~for
scanning thresholds of particle production. Furthermore, polarisation of
electrons and positrons is possible.
In proton collisions the rate of unwanted collision processes is very high,
whereas the pointlike nature of leptons results in low backgrounds.
In addition, a linear collider offers besides $e^+e^-$ collisions the
options of $e^-e^-$, $e\gamma$ and $\gamma\gamma$ collisions which could
provide important additional insight. \\
Hadron and Lepton Colliders are complementary and the present state of
knowledge in particle physics would not have been achieved without both
types of colliders running concurrently.

Telling examples from the past are internal consistency tests of the
electroweak part of the Standard Model. In 1994, the precision electroweak
measurements of the $Z^0$ boson predicted a mass of the top quark
from quantum corrections
of $\mathrm {M_{top}}$ = 178$\pm $11 $^{+18}_{-19}$ GeV. The direct
measurement at the Tevatron in the following years yielded 
$\mathrm {M_{top}}$ = 174.3 $\pm 5.1$ GeV.
The indirect measurement of $\mathrm {M_W}$ = 80.363 $\pm$ 0.032 GeV 
agrees well with the direct mass measurements from Tevatron and LEP of
$\mathrm {M_{W}}$ = 80.450 $\pm$ 0.039 GeV. The Standard Model has been
tested and so far confirmed at the quantum level. 

\begin{figure}[h]
\epsfxsize200pt
\figurebox{200pt}{160pt}{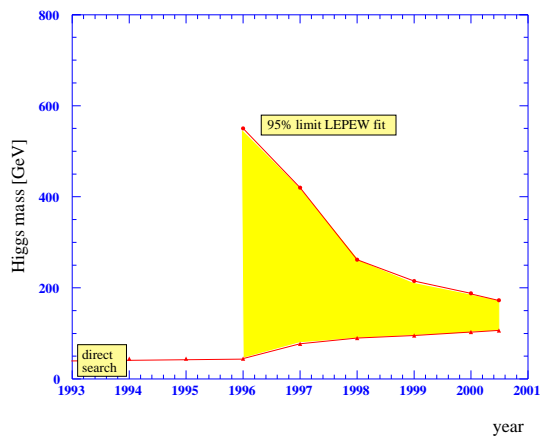}
\caption{Development over the past years of limits for the mass of the 
Higgs boson from direct and indirect measurements.}
\label{fig:historymh}
\end{figure}

Much progress about the possible mass range of the Higgs boson, if it exists,
has been achieved in the past around five years. Lower bounds on the mass
have been derived through direct searches at LEP running with ever increasing
centre-of-mass energies until the year 2000. The Standard Model Higgs 
contribution to electroweak observables through loop corrections
provides further indirect information.
Although these corrections vary only logarithmically, $\propto$ log($M_H/M_W$),
the accuracy of the electroweak data obtained at LEP, SLC and the Tevatron,
provides sensitivity to $M_H$ and in turn an upper bound for the allowed
mass range. The development of these bounds is shown in 
figure~\ref{fig:historymh}. The 95\% upper limit for $M_H$ of 196 GeV
is well within the reach of a linear collider with a centre-of-mass energy
of 500 GeV. \\

In general, the physics target of the next generation of 
electron-positron linear colliders
will be a comprehensive and high precision coverage of the energy range
from $M_Z$ up to around 1 TeV.
Energies up to around 3 to 5 TeV could be achieved with the 
following generation of colliders. The physics case for such a machine
will depend on the results from the LHC and the linear collider
in the sub-TeV range.

\section{Selected Physics Topics}

In this chapter, some of the main physics topics to be studied
at a linear collider will be discussed. Emphasis is given to the study
of the Higgs mechanism in the Standard Model, the measurements
of properties of supersymmetric particles, and precision tests of 
the electroweak theory.
More details about these topics as well as information about the
numerous topics not presented here can be found in the physics
books published in the studies of the physics potential of future
linear colliders~\cite{ecfaphys,snowmass,jlcphys,fnal}. 

\subsection{Standard Model Higgs Boson}\label{subsec:higgs}

The main task of a linear electron-positron collider will be to
establish experimentally the Higgs mechanism as the mechanism for 
generating the masses of fundamental particles:
\begin{itemize}
\item The Higgs boson must be discovered.
\item The couplings of the Higgs boson to gauge bosons and to fermions
must be proven to increase with their masses.
\item The Higgs potential which generates the non-zero field in the vacuum must
be reconstructed by determining the Higgs self-coupling.
\item The quantum numbers ($J^{PC}=0^{++}$) must be confirmed.
\end{itemize}

\begin{figure}
\epsfxsize200pt
\figurebox{}{200pt}{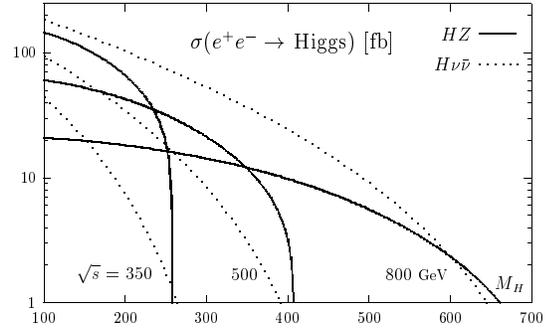}
\caption{The Higgs-strahlung and WW fusion production cross-sections as a
function of $M_H$ for different $\sqrt{s}$.}
\label{fig:hcrosssect}
\end{figure}

The main production mechanisms for Higgs bosons in $e^+e^-$ collisions
are Higgs-strahlung $e^+e^- \rightarrow HZ$ and WW-fusion
$e^+e^- \rightarrow \nu_e \bar \nu_e H$, and 
the corresponding cross-sections as a function
of $M_H$ are depicted in figure~\ref{fig:hcrosssect} for three different
centre of mass energies.
With an integrated luminosity of 500 $fb^{-1}$, corresponding to about two
years of operation, some $10^5$ events will be produced
and can be selected with high efficiency and very low background. 

\begin{figure}[h]
\epsfxsize200pt
\figurebox{}{200pt}{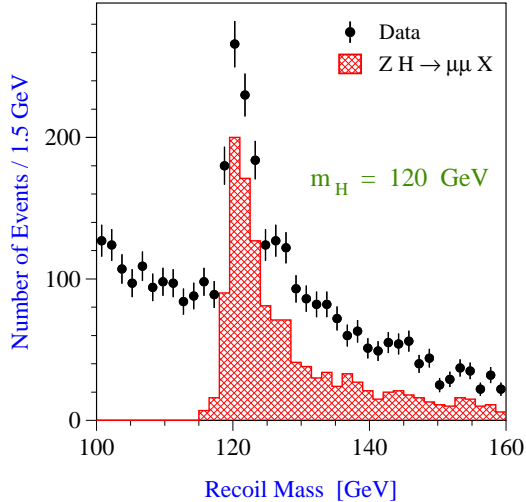}
\caption{The $\mu^+ \mu^-$ recoil mass distribution in the process 
$e^+e^- \rightarrow HZ \rightarrow {\mu}^+{\mu}^-$ for $M_H$=120 GeV, 
500 $fb^{-1}$ at $\sqrt{s}$=350 GeV}. 
\label{fig:hmass}
\end{figure}

The Higgs-strahlung
process $e^+e^- \rightarrow ZH$, with $Z \rightarrow {\ell}^+{\ell}^-$, offers
a very distinctive signature ensuring the observation of the Standard
Model Higgs boson up to the kinematical limit independently of its decay
as illustrated in figure~\ref{fig:hmass}. 

\begin{figure}[h]
\epsfxsize200pt
\figurebox{200pt}{200pt}{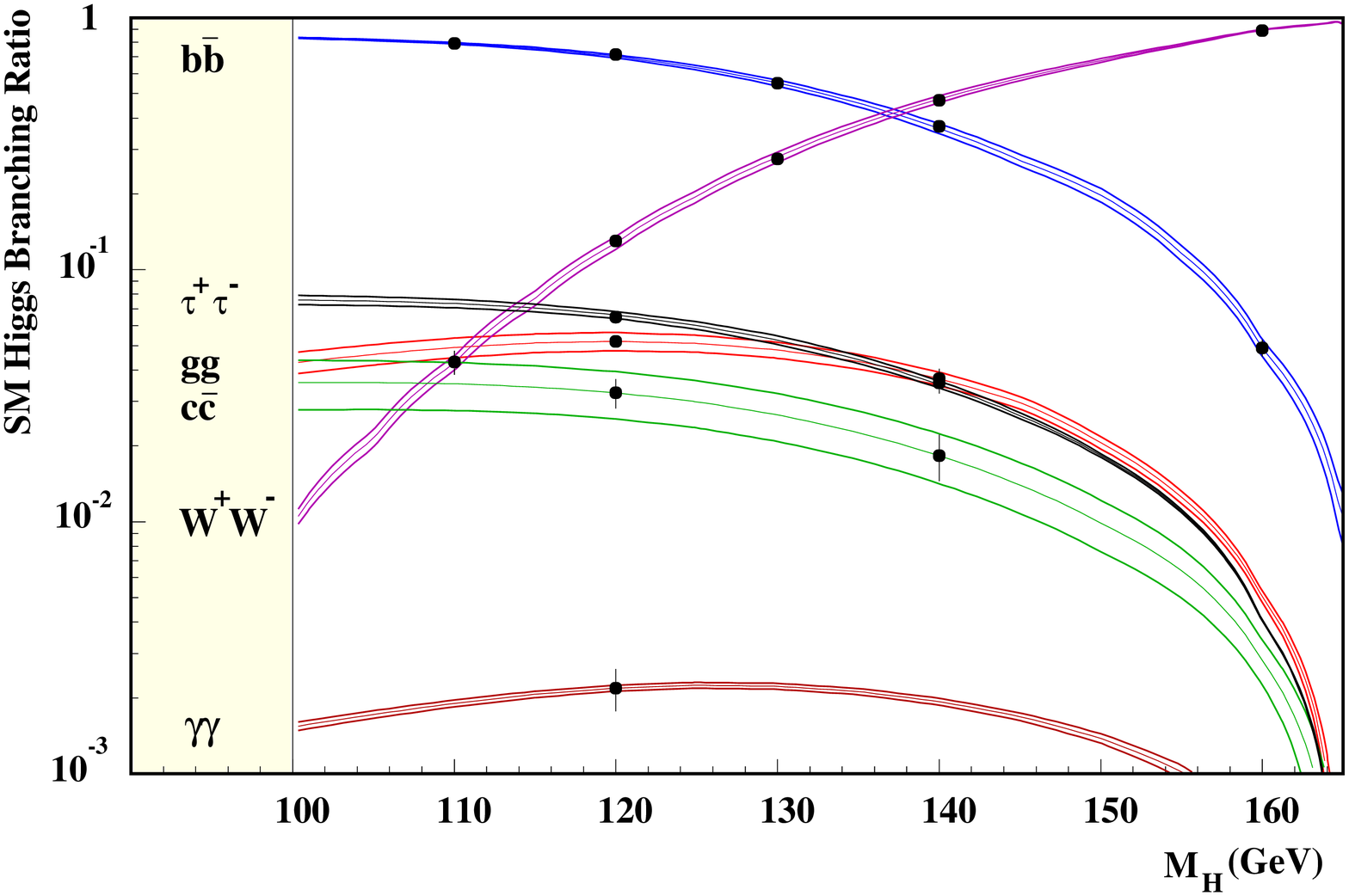}
\caption{The predicted Standard Model Higgs boson branching ratios (bands)
and the expected experimental accuracies (points with error bars).}
\label{fig:higgsbr}
\end{figure}

The Higgs-strahlung process allows to measure the decay branching ratios
of the Higgs boson and to test their dependence on the mass of the 
fundamental particles.
The detectors proposed for linear colliders have excellent flavour tagging
capability in order to distinguish the different hadronic decay modes
(see for example\cite{det}). Therefore,
the branching ratios can be determined with accuracies of a few percent,
as shown in figure~\ref{fig:higgsbr}.

The determination of the Yukawa
coupling of the Higgs boson to the top quark is provided by the
process $e^+e^- \rightarrow t\bar t H$ at $\sqrt{s}$ of
about 800 GeV; for 1000 $fb^{-1}$ an accuracy of 6\% can be expected. \\
The Higgs boson quantum numbers can be determined through the rise of the 
cross section close to the production threshold and through the angular
distributions of the $H$ and $Z$ bosons in the continuum.

\begin{figure}[htb]
\epsfxsize200pt
\figurebox{}{}{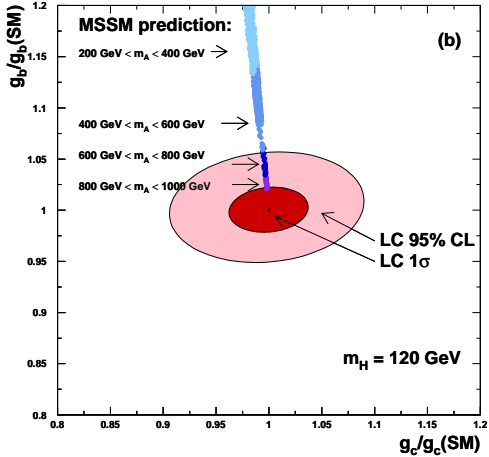}
\caption{Higgs coupling determination: The contours for $g_b$ vs. $g_c$
for a 120 GeV Higgs boson normalised to their Standard Model expectations
as measured with 500$fb^{-1}$.}
\label{fig:higgscoupl}
\end{figure}

The Higgs boson production and decay rates discussed above, can be used
to determine the Higgs couplings to gauge bosons and fermions. A global fit
to the measured observables optimises the available information, accounts
properly for the experimental correlations between the different 
measurements and allows to  extract the Higgs couplings in a model
independent way. As an example for the accuracies reachable with the 
newly developed program HFITTER~\cite{ecfaphys}, 
figure~\ref{fig:higgscoupl} shows 
1$\sigma$ and 95\% confidence level contours for the fitted values of the
couplings $g_c$ and $g_b$ to the charm and bottom quark with comparison
to the sizes of changes expected from the minimal supersymmetric
extension to the Standard Model (MSSM).

\begin{figure}[hbt]
\epsfxsize200pt
\figurebox{}{}{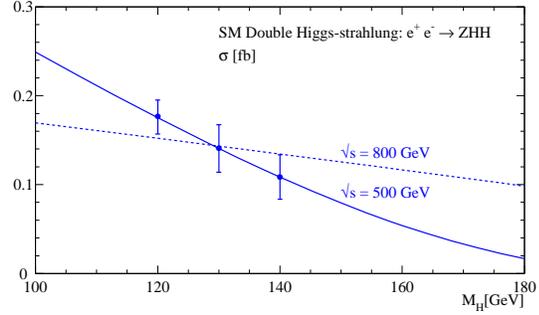}
\caption{The cross-section for double Higgs-strahlung in the Standard Model
at $\sqrt{s}$ of 500 and 800 GeV together with the experimental accuracies 
expected for 1 $ab^{-1}$ (points with error bars).}
\label{fig:hhz}
\end{figure}

To generate a non-zero value of the Higgs field in the vacuum, the minimum
$\mathrm {\phi}_0 = v/\sqrt{2}$
of the self potential of the Higgs field
$\mathrm V = \lambda{({\phi}^2 - \frac{1}{2} {v}^2)}^2$
must be shifted away from the origin. This potential can be reconstructed
by measuring the self couplings of the physical Higgs boson as predicted by 
the potential $\mathrm V = \lambda v^2 H^2 + \lambda v H^3 + 
\frac{1}{4} \lambda H^4$. The trilinear Higgs coupling 
$\lambda_{HHH} = 6\lambda v$
can be measured directly in the double
Higgs-strahlung process $e^+e^- \rightarrow HHZ \rightarrow q\bar q b\bar b
b\bar b$. The final state contains six partons resulting in a rather
complicated experimental signature with six jets, a challenging task
calling for excellent granularity of the tracking device and the
calorimeter~\cite{det}.
Despite the low cross section of the order of  
0.2 $fb$ for $M_H$ = 120 GeV at $\sqrt{s}$ = 500 GeV,
the coupling can be measured with an accuracy of better
than 20\% for Higgs masses below 140 GeV at $\sqrt{s}$=500 GeV with an
integrated luminosity of 1$ab^{-1}$ as shown in 
figure~\ref{fig:hhz}.

Measurements of Higgs boson properties and their anticipated accuracies
are summarised in table~\ref{tab:higgs}.

\begin{table}
\begin{center}
\caption{Precision of the possible measurements of the Higgs boson
properties.}
\label{tab:higgs}
\begin{tabular}{|l|c|c|}\hline
$m_H$             & 120 GeV & 140 GeV \\ \hline
mass            & 0.06\% & 0.05\%  \\
spin            & yes      & yes     \\
CP              & yes      & yes   \\ \hline
total width     & 6 \%  & 5 \% \\ \hline
$g_{HZZ}$       & 1 \%  &  1 \%  \\
$g_{HWW}$       & 1 \%  &  2 \% \\
$g_{Hbb}$       & 2 \%  &  2 \% \\
$g_{Hcc}$       & 3 \%  & 10 \% \\
$g_{H\tau\tau}$ & 3 \%  &  5 \% \\
$g_{Htt}$       & 3 \%  &  6 \% \\
$\lambda_{HHH}$ & 20\% & $\sim$30\% \\ \hline
\end{tabular}
\end{center}
\end{table}

In summary, the Higgs mechanism can be established in an unambiguous way 
at a high luminosity electron-positron collider with a centre-of-mass
energy up to around one TeV as the mechanism responsible for the spontaneous
symmetry breaking of the electroweak interactions.

\subsection{Supersymmetric Particles}\label{subsec:susy}

Supersymmetry (SUSY) is considered the most attractive 
extension of the Standard Model, which cannot be the ultimate
theory for many reasons. The most important feature of SUSY is that it
can explain the hierarchy between the electroweak scale of $\approx$100 GeV,
responsible for the W and Z masses, and the Planck scale 
$M_{Pl} \simeq 10^{19}$ GeV.
When embedded in a grand-unified theory, it makes a very precise 
prediction of the electroweak mixing angle 
$\sin^2{\theta_W}$ in excellent concordance with the
precision electroweak measurement. 
In the following, only the minimal supersymmetric extension to the
Standard Model (MSSM) will be considered and measurements of the properties
of the supersymmetric particles will be discussed. Studies of the
supersymmetric Higgs sector can be found 
elsewhere\cite{ecfaphys,snowmass,jlcphys,fnal}. \\

In addition to the particles of the Standard Model, the MSSM contains their
supersymmetric partners: sleptons $\tilde{l}^{\pm}, \tilde{\nu}_l$ 
($l = e, \mu, \tau$), squarks $\tilde{q}$,
and gauginos $\tilde{g}, \tilde{\chi}^{\pm}, \tilde{\chi}^0$. 
In the MSSM the multiplicative quantum number R-parity is conserved,
$R_p = +1$ for particles and $R_p = -1$ for sparticles.
Sparticles are therefore produced in pairs
and they eventually decay into the lightest 
sparticle which has to be stable. As an example, smuons are produced and
decay through the process 
$e^+e^- \rightarrow \tilde{\mu}^+ \tilde{\mu}^- \rightarrow \mu^+ \mu^- 
\chi^0_1 \chi^0_1$ with $\chi^0_1$ as the lightest sparticle being stable
and, therefore, escaping detection. \\

\begin{figure}
\epsfxsize200pt
\figurebox{}{}{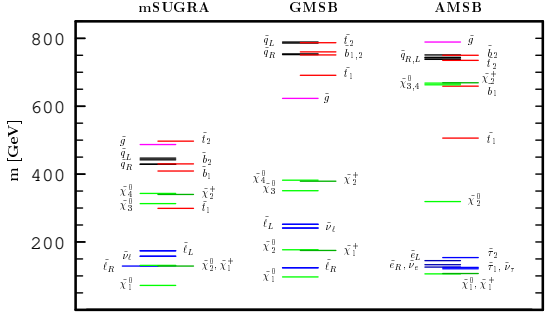}
\caption{Examples of mass spectra in mSUGRA, GMSB and AMSB models.}
\label{fig:susyspectra}
\end{figure}

The mass scale of sparticles is only vaguely known. In most scenarios
some sparticles, in particular charginos and neutralinos, are expected to
lie in the energy region accessible by the next generation of 
$e^+e^-$ colliders also supported by the recent measurement of 
$(g-2)_{\mu}$~\cite{g2}. 
Examples of mass spectra for three SUSY breaking
mechanisms (mSUGRA, GMSB, AMSB) are given in figure~\ref{fig:susyspectra}. \\
 
The most fundamental problem of supersymmetric theories is how SUSY
is broken and in which way this breaking is communicated to the
particles. Several scenarios have been proposed in which the mass spectra are
generally quite different as illustrated in figure~\ref{fig:susyspectra}. 
High  
precision measurements of the particle properties are therefore expected 
to distinguish between some of these scenarios. The study and exploration of
Supersymmetry will proceed in the following steps:
\begin{itemize}
\item Reconstruction of the kinematically accessible spectrum of 
sparticles and the measurement of their properties, masses and
quantum numbers 
\item Extraction of the basic low-energy parameters such as mass parameters,
couplings, and mixings
\item Analysis of the breaking mechanism and reconstruction of the
underlying theory.
\end{itemize}

\begin{figure}[htb]
\epsfxsize200pt
\figurebox{}{}{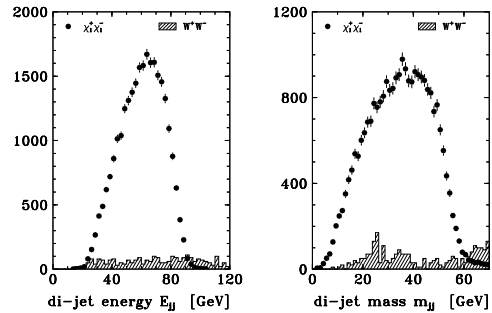}
\caption{Di-jet energy and mass spectra at $\sqrt{s}$=320 GeV and 160 $fb^{-1}$
for the reaction $e^+e^- \rightarrow \tilde{\chi}^+_1 \tilde{\chi}^-_1
\rightarrow {\ell}^{\pm} \nu \tilde{\chi}^0_1 q\bar q' \tilde{\chi}^0_1$.}
\label{fig:chmassdir}
\end{figure}

\begin{figure}[hbt]
\epsfxsize200pt 
\begin{sideways}
\figurebox{}{}{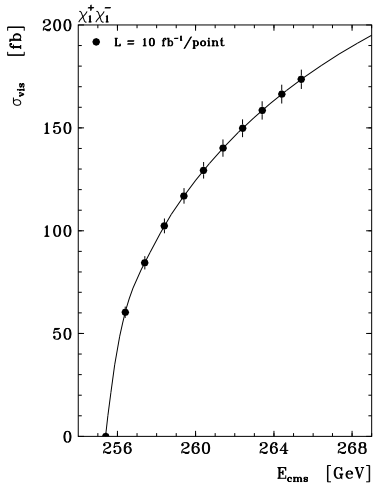}
\end{sideways}
\caption{Cross section near threshold for the process
$e^+e^- \rightarrow \tilde{\chi}^+_1 \tilde{\chi}^-_1$,
10 $fb^{-1}$ per point.}
\label{fig:chmassscan}
\end{figure}

While it is unlikely that the complete spectrum of sparticles will be
accessible at a collider with $\sqrt{s}$ up to around 1 TeV, a significant
part of the spectrum should be measureable.
In general, at an $e^+e^-$ collider production cross sections are large
and backgrounds are rather small.
Masses of sparticles can be determined from the decay kinematics,
measured in the continuum. 
An example for such mesurements is given in figure~\ref{fig:chmassdir}.
Typical accuracies are of the order 100 to 300 MeV.
Excellent mass resolutions of the order of 50 MeV with an integrated
luminosity of 100 $fb^{-1}$ can be obtained for the light charginos and
neutralinos through the measurement of the excitation curves at 
production threshold, as also shown in figure~\ref{fig:chmassscan}. 


The reconstruction of the mechanism which breaks supersymmetry will give
significant insight into the laws of Nature at energy scales where gravity
becomes important as quantum effect. Various models like minimal supergravity 
mSURGA), gauge mediated SUSY breaking (GMSB), or anomaly mediated
SUSY breaking (AMSB) have been proposed. These mechanisms lead to different
spectra of sparticle masses as was shown already in 
figure~\ref{fig:susyspectra}. The supersymmetric renormalisation group
equations (RGE's) are largely independent of the assumed properties of the
specific SUSY model at high energies. This can be used to interpret
the measured SUSY spectra. In a 'bottom up' approach, the measured
electroweak scale SUSY parameters are extrapolated to high energies using
these RGE's. 

\begin{figure*}[hbt]
\epsfxsize200pt
\includegraphics[width=0.65\linewidth,clip]
{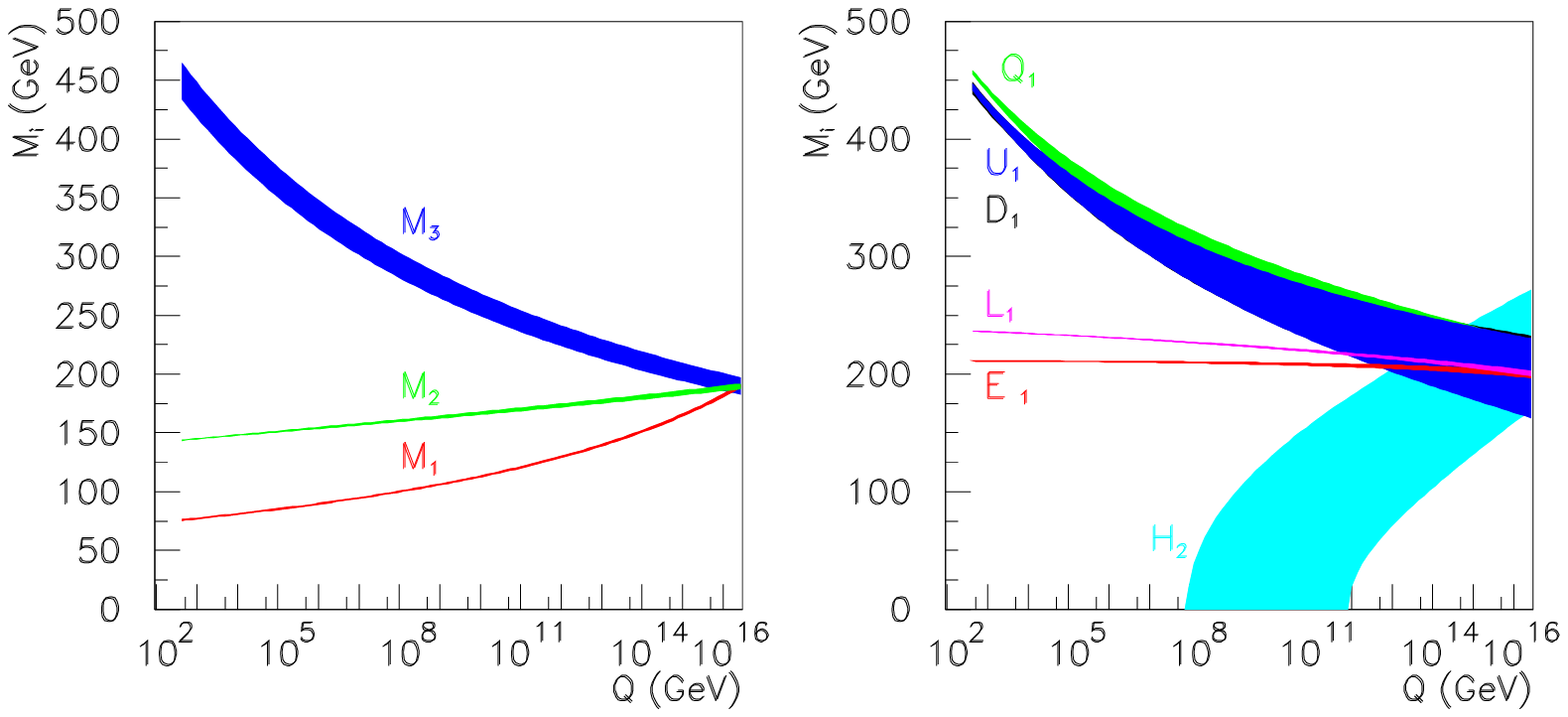} \hspace{-5mm}
\epsfxsize200pt 
\includegraphics[width=0.35\linewidth,clip]
{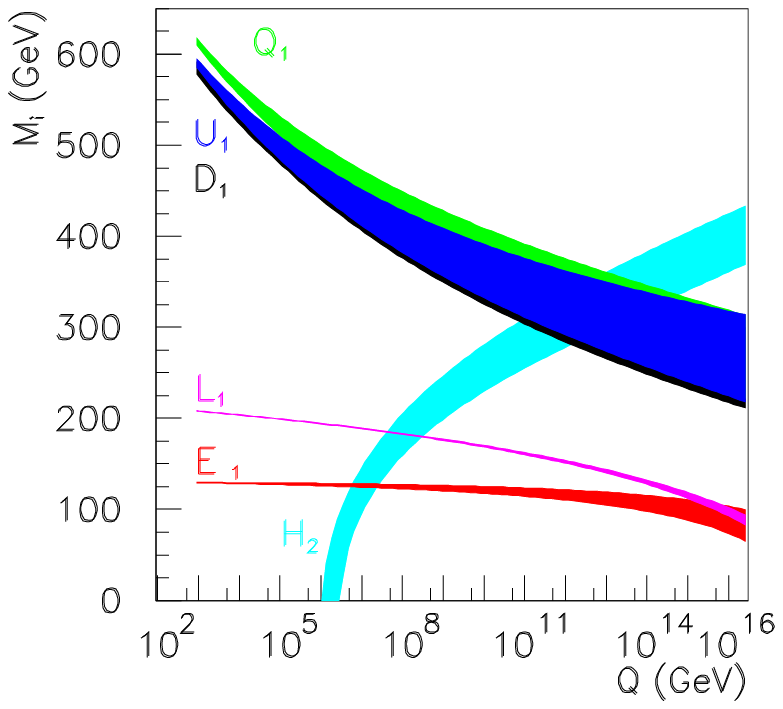}
\caption{Extrapolation of SUSY parameters measured at the electroweak scale
to high energies.}
\label{fig:extrapol}
\end{figure*}

Due to the high precision of the measured input variables, only possible
at the linear collider, an accurate test can be performed at which energy
scale certain parameters become equal. Most interesting,
the assumption of grand unification of forces requires the gaugino mass
parameters $M_1, M_2, M_3$ to meet at the GUT scale
(figure~\ref{fig:extrapol} (left)). 
Different SUSY breaking mechanisms predict different unification patterns
of the sfermion mass parameters at high energy. With the high accuracy of
the linear collider measurements these models can be distinguished as 
shown in figure~\ref{fig:extrapol} for the case of mSUGRA (middle) 
and GMSB (right). \\

In summary, the high precision studies of supersymmetric particles and
their properties can open a window to energy scales far above the scales
reachable with future accelerators, possibly towards the
Planck scale where gravity becomes important.

\subsection{Precision Measurements}\label{subsec:prectest}

\begin{figure*}
\epsfxsize200pt
\figurebox{}{100pt}{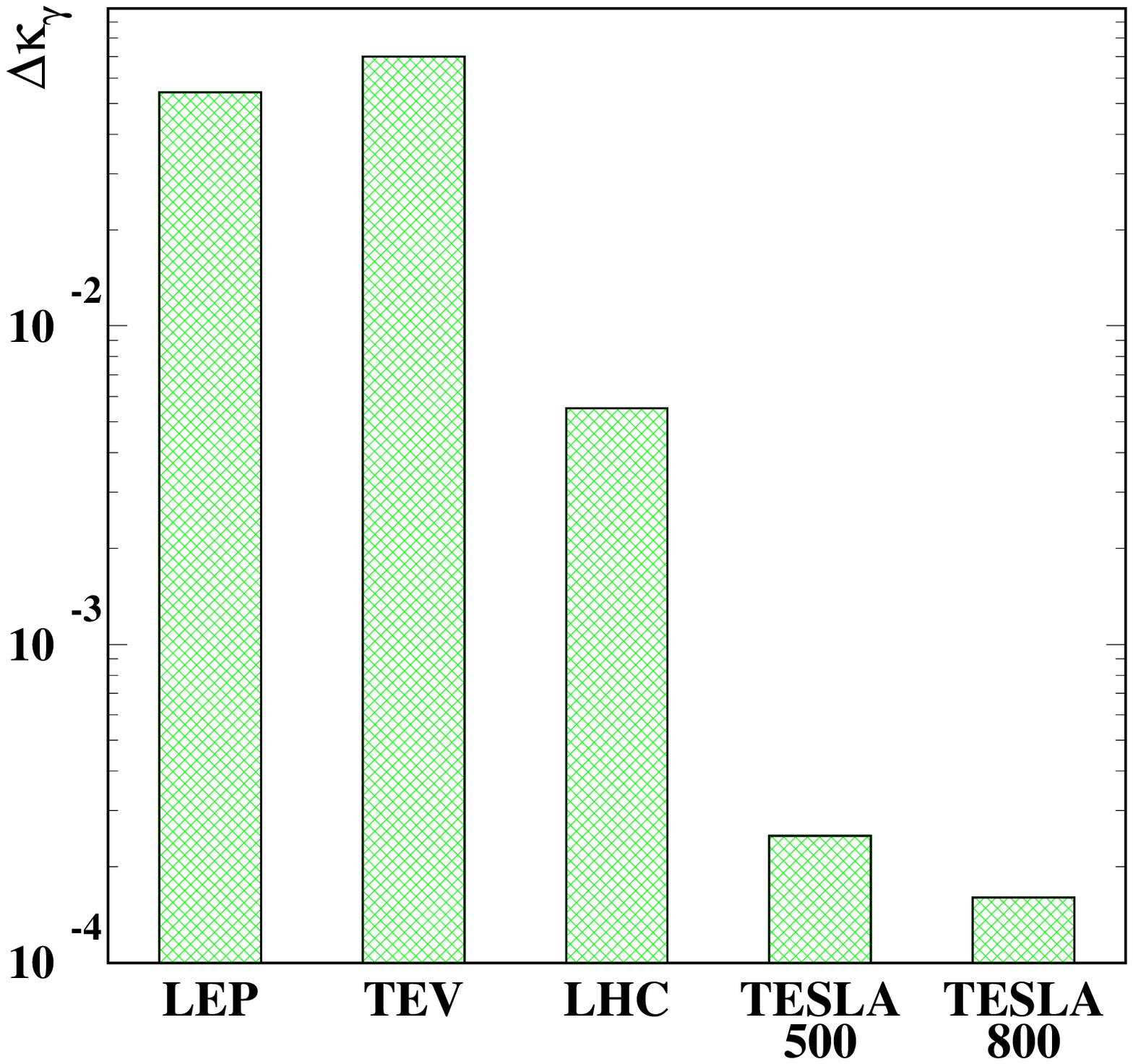}
\epsfxsize200pt
\figurebox{}{200pt}{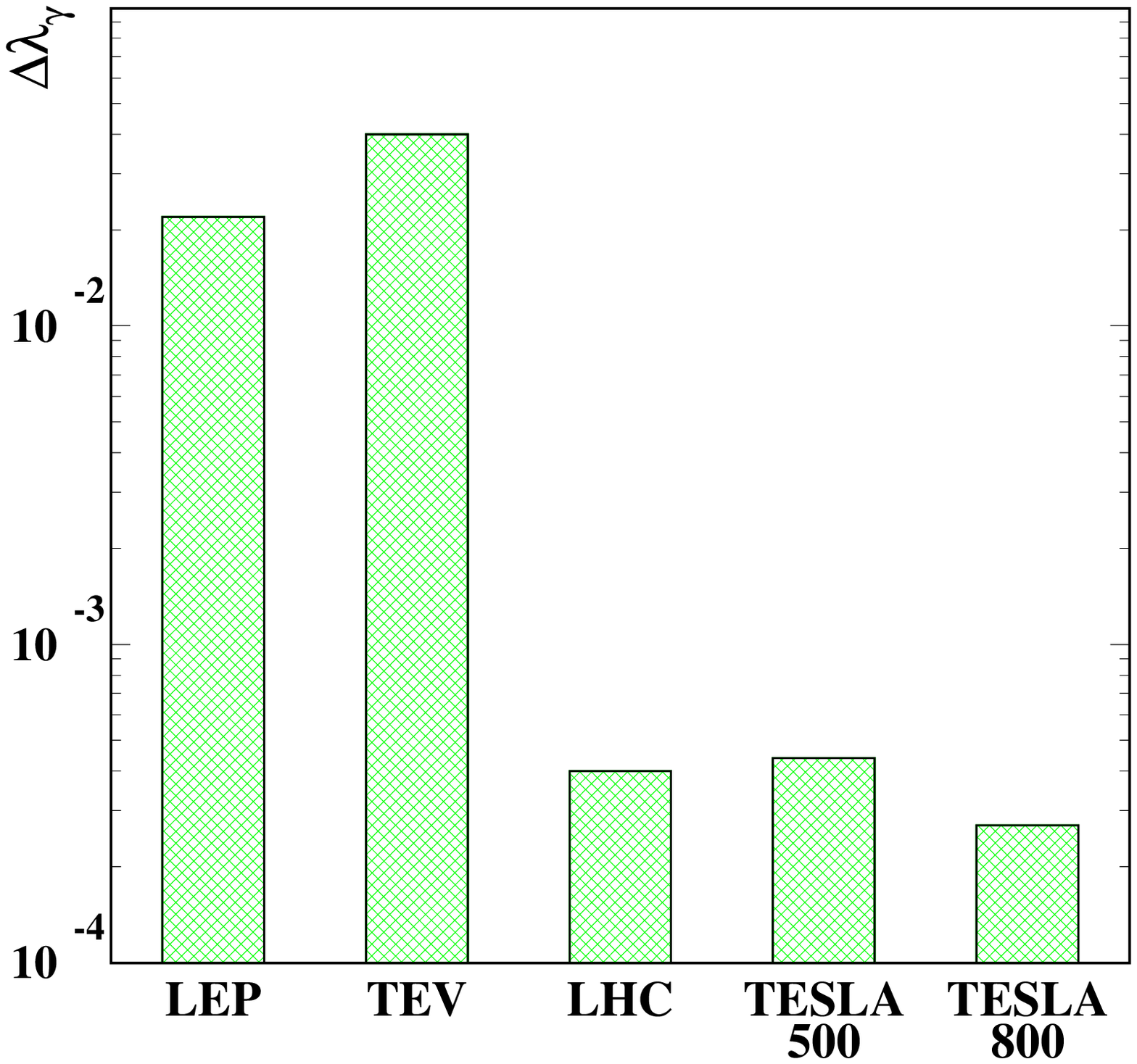}
\caption{Comparison of constraints on the anomalous couplings 
$\Delta \kappa_{\gamma}$ and $\Delta \lambda_{\gamma}$ at different
machines}
\label{fig:tgc}
\end{figure*}

The primary goal of precision measurements of gauge boson properties 
is to establish the non-abelian nature of electroweak interactions.
The gauge symmetries of the Standard Model determine the form and the
strength of the self-interactions of the electroweak bosons, the
triple couplings $WW\gamma$ and $WWZ$ and the quartic couplings. 
Deviations from the Standard Model expectations for these couplings
could be expected in
several scenarios, for example in models where there exists no light
Higgs boson and where the $W$ and $Z$ bosons are generated dynamically and
interact strongly at high scales. Also for the
extrapolation of couplings to high scales to test theories of grand
unification such high precision measurements are mandatory.
For the study of the couplings between gauge bosons  the best precision 
is reached at the highest possible
centre of mass energies. These
couplings are especially sensitive to models of strong electroweak
symmetry breaking. \\
W bosons are produced either in pairs, $e^+e^- \rightarrow W^+W^-$ or
singly, $e^+e^- \rightarrow We \nu$ with both processes being sensitive 
to the triple gauge couplings. In general the total errors estimated on the
anomalous couplings are in the range of few $\times 10^{-4}$.
Figure~\ref{fig:tgc} compares the precision obtainable for 
$\Delta \kappa_{\gamma}$ and $\Delta \lambda_{\gamma}$ at different
machines.

The measurements at a linear collider are sensitive to strong symmetry
breaking beyond $\Lambda$ of the order of 5 TeV, to be compared
with the electroweak symmetry breaking scale $\Lambda_{EWSB} = 4 \pi v 
\approx$ 3 TeV. \\
  
\begin{figure}
\epsfxsize200pt
\figurebox{}{200pt}{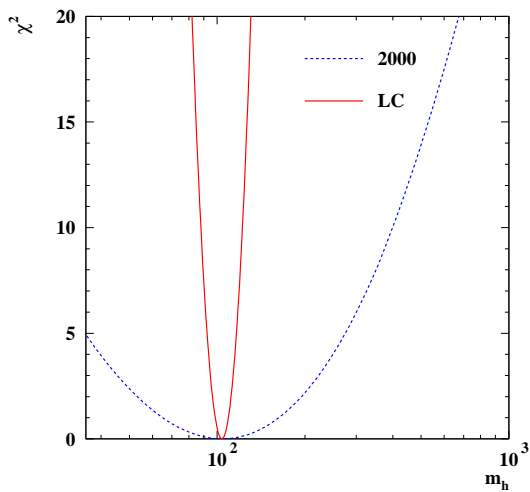}
\caption{$\Delta{\chi}^2$ as a function of the Higgs boson mass for the
electroweak precision data today (2000) and after GigaZ running (LC).}
\label{fig:gigaz}
\end{figure}

One of the most sensitive quantities to loop corrections from the
Higgs boson is the effective weak mixing angle in $Z$ boson decays. 
By operating the collider at energies close to the $Z$-pole
with high luminosity (GigaZ) to collect at least $10^9$ $Z$ bosons
in particular the accuracy of the measurement of $\sin^2{\theta}^l_{eff}$ 
can be improved by one order of magnitude wrt. the precision obtained
today~\cite{drees}. With both electron and positron beams longitudinally
polarised, $\sin^2{\theta}^l_{eff}$ can be determined most accurately by
measuring the left-right asymmetry $A_{LR} = A_e =
2 v_e a_e/(v_e^2 + a_e^2)$ with $v_e$ ($a_e$) being the vector 
(axialvector)
couplings of the $Z$ boson to the electron and $v_e/a_e$ = 1 - 
$4 \sin^2{\theta}^l_{eff}$
for pure $Z$ exchange.
Particularly demanding is the precision of $2 \times 10^{-4}$ with which the 
polarisation needs to be known to match the statistical accuracy.
An error in the weak mixing angle of $\Delta \sin^2{\theta}^l_{eff}$ =
0.000013 can be expected.  Together with an improved determination
of the  mass of the $W$ boson to a precision of some 6 MeV through a
scan of the $WW$ production threshold and with the measurements obtained at
high energy running of the collider this will allow many high precision
tests of the Standard Model at the loop level. As an example, 
figure~\ref{fig:gigaz} shows the variation of the fit ${\chi}^2$ to the
electroweak measurements as a function of $M_H$ for the present data
and for the data expected at a linear collider. The mass of the Higgs 
boson can indirectly be constraint at a level of 5\%. Comparing this
prediction with the direct measurement of $M_H$ consistency tests of the
Standard Model can be performed at the quantum level or to measure free
parameters in extensions of the Standard Model.
This is of particular importance if $M_H > 200$ GeV in contradiction to
the current electroweak measurements. \\

In summary, there is strong evidence for new phenomena at the TeV energy scale.
Only the precision exploration at the linear collider will allow, together
with the results obtained at the Large Hadron Collider, the understanding of
the underlying physics and will open a new window beyond the centre-of-mass
energies reachable. Whatever scenario is realized in nature, the linear
collider will add crucial information beyond the LHC. 
There is global consensus in the high energy physics community that the
next accelerator based project needs to be an electron-positron linear
collider with a centre-of-mass energy of at least 500 GeV.

\section{Electron-Positron Linear Colliders}

The feasibility of a linear collider has been successfully demonstrated by the
operation of the SLAC Linear Collider, SLC. However, aiming at centre-of-mass
energies at the TeV scale with luminosities of the order of 
$10^{34} cm^{-2}s^{-1}$ 
requires at least two orders of magnitude higher beam power and
two orders of magnitude smaller beam sizes at the interaction point.
Over the past decade, several groups worldwide have been pursuing different
linear collider designs for the centre-of-mass energy range up to around
one TeV as well as for the multi-TeV range.
Excellent progress has been achieved at various test facilities worldwide
in international collaborations on crucial aspects of the collider designs.
At the Accelerator Test Facility at KEK~\cite{ATF}, emittances within a 
factor two of the damping ring design have been achieved.
At the Final Focus Test Beam at SLAC~\cite{FFTB} demagnification of the
beams has been proven; the measured spot sizes are well in agreement with
the theoretically expected values. 
The commissioning and operation of the TESLA Test Facility at 
DESY~\cite{TTF} has
demonstrated the feasibility of the TESLA technology.
In the following, a short review of the different approaches is given.

\subsection{TeV range} 

Three design studies are presently pursued: JLC~\cite{jlc}, NLC~\cite{nlc} 
and TESLA~\cite{tesla}, centred around KEK, SLAC and DESY, respectively.
Details about the design, the status of development and the individual
test facilities can be found in the above quoted references as well as 
in the status reports presented at LCWS2000~\cite{napoly,raub,chin}. 
A comprehensive summary of the present
status can be found in the Snowmass Accelerator R\&D Report~\cite{snowacc},
here only a short discussion of the main features and 
differences of the
three approaches will be given with emphasis on luminosity and energy
reach. \\
One key parameter for performing the physics program at a collider
is the centre-of-mass energy achievable. 
The energy reach of a collider with a given linac length and
a certain cavity filling factor is determined
by the gradient achievable with the cavity technology chosen.
For normalconducting cavities the maximum achievable gradient scales
roughly proportional to the RF frequency used, for superconducting
Niobium cavities, the fundamental limit today is around 55 MV/m. \\
The second key parameter
for the physics program is the luminosity $\cal{L}$, given by
\begin{equation}
{\cal{L}} = \frac{n_b N^2_e f_{rep}}{4 \pi \sigma^{\ast}_x 
\sigma^{\ast}_y} H_D
\label{eq:lumi1} 
\end{equation}
where $n_b$ is the number of bunches per pulse containing $N_e$ particles,
$f_{rep}$ is the pulse repetition frequency, $\sigma^{\ast}_{x,y}$ 
the horizontal (vertical) beamsize at the interaction point, and $H_D$
the disruption enhancement factor.
An important constraint on the choice of the parameters is the effect
of beamstrahlung due to the emission of synchrotron radiation. The
average fractional beam energy loss $\delta_E$ is proportional to
$\frac{1}{(\sigma^{\ast}_x + \sigma^{\ast}_y)^2}$. 
Choosing a flat beam size ($\sigma^{\ast}_x \gg \sigma^{\ast}_y$) at the
interaction point, $\delta_E$ becomes independent of the vertical beam size
and the luminosity can be increased by reducing $\sigma^{\ast}_y$ as much as
possible. Since $\sigma^{\ast}_y \propto \sqrt{\epsilon_{y} \beta^{\ast}_y}$
this can be achieved by a small vertical beta function $\beta^{\ast}_y$ 
and a small normalised vertical emittance $\epsilon_{y}$ at the 
interaction point.  
The average beam power $P_{beam} = \sqrt{s} n_b N_e f_{rep}
=\eta P_{AC}$ is obtained from the mains power $P_{AC}$ with an
efficiency $\eta$. Equation~(\ref{eq:lumi1}) can then be rewritten as
\begin{equation}
{\cal{L}} \propto \frac{\eta P_{AC}}{\sqrt{s}} \sqrt{\frac{\delta_E}
{\epsilon_y}}
\end{equation}
High luminosity
therefore requires high efficiency $\eta$ and high beam quality with low
emittance $\epsilon_{y}$ and low emittance dilution 
$\Delta \epsilon/\epsilon \propto f^6_{RF}$, which is largely determined 
by the RF frequency $f_{RF}$ of the chosen technology.

\begin{figure}[htb]
\epsfxsize200pt
\figurebox{}{400pt}{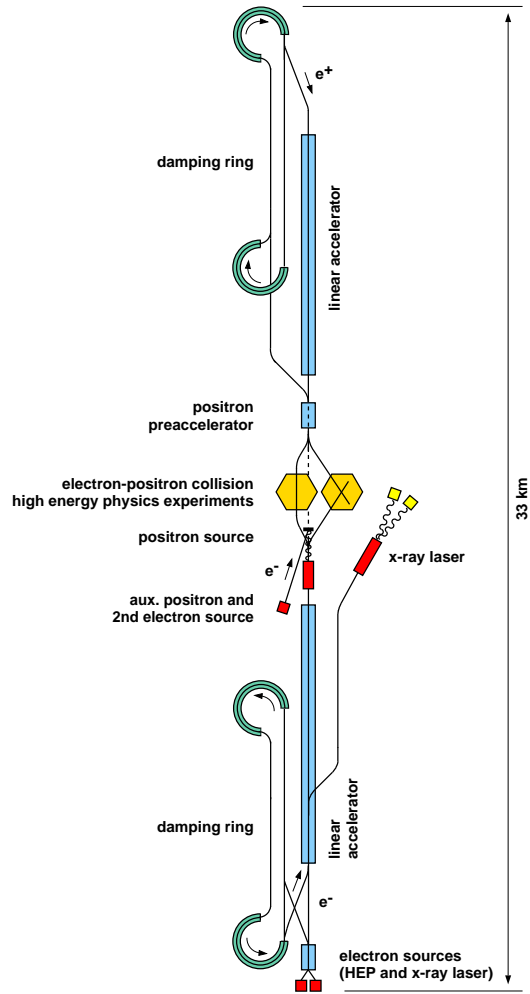}
\caption{Schematic layout of TESLA}
\label{fig:tesla}
\end{figure}

The fundamental difference between the three designs
is the choice of technology for the accelerating structures.   
The design of NLC is based on normalconducting cavities using
$f_{RF}$ of 11.4 GHz (X-band), for JLC two options, 
X-band or C-band (5.7 GHz) are pursued.
The TESLA concept, developed by the TESLA collaboration,
is using superconducting cavities (1.3 GHz). 
As an example for a linear collider facility,
figure~\ref{fig:tesla} shows the schematic layout of TESLA.

Table~\ref{tab:comp} compares some key parameters for the different 
technologies at $\sqrt{s}$ = 500 GeV,
like repetition rate $f_{rep}$ for bunch trains with $N_b$
bunches, the time $\Delta T_b$ between bunches within a train which
allows head on crossing of the bunches for TESLA but requires a crossing
angle for the other designs. The design luminosity ${\cal{L}}$, 
beam power $P_{beam}$
and the required mains power $P_{AC}$ illustrate that for a given mains power
the superconducting technology delivers higher luminosity. On the other hand
the lower gradient $G_{acc}$ requires a longer linac for the same 
centre-of-mass energy reach. 
As can be seen from table~\ref{tab:comp} the X-band machines
call for a beam loaded
(unloaded) gradient of some 50 (70) MV/m for $\sqrt{s}$ of 500 GeV.
Recently, it has been found that high gradient operation of normalconducting
cavities results in surface damage of the structures. 
Intense R\&D is going on
in collaboration between SLAC, KEK and CERN in order to understand and
resolve the problem. At present it seems that the onset of the damage
depends on the structure length and the group velocity within the
cavity~\cite{nlc}.

\begin{figure}
\epsfxsize200pt
\figurebox{}{200pt}{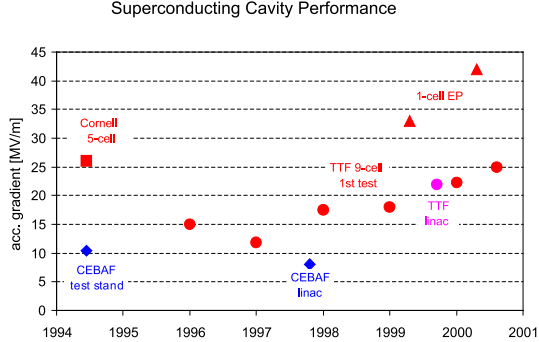}
\caption{Evolution of superconducting cavity performance. The 
average gradient achieved with TESLA 9-cell cavities produced
in industry (first test, no additional processing) is shown as dots.}
\label{fig:cavperform}
\end{figure}

The TESLA design requires 23.5 MV/m for $\sqrt{s}$ = 500 GeV, 
a gradient which is
meanwhile routinely achieved for cavities fabricated in industry
as illustrated in figure~\ref{fig:cavperform}.

\begin{table*}
\begin{center}
\caption{Comparison of some crucial parameters at 500 GeV for the different
technologies under study, see text for details.}\label{tab:comp} 
\begin{tabular}{|l|l|l|l|l|}
   \hline
    & TESLA & NLC & JLC-X & JLC-C \\ 
\hline
$f_{rep}$ [Hz] & 5 & 120 & 150 & 100 \\
$N_b$ & 2820 & 190 & 190 & 142  \\
$\Delta T_b$ [ns] & 337 & 1.4 & 1.4 & 2.8 \\
bunch crossing & head on & angle & angle & angle \\
$N_e/bunch [10^{10}]$ & 2 & .75 & 0.7 & 1.11 \\ \hline 
$\sigma^*_{x/y} [nm]$ & 553/5  & 245/2.7 & 239/2.57 & 318/4.3 \\
$\delta_E [\%]$ & 3.2 & 4.7 & 5.3 & 3.9 \\ \hline 
${\cal L} [10^{34}cm^{-2}s^{-1}]$ & 3.4 & 2 & 2.64 & 1.3 \\ \hline
$P_{beam} [MW]$ & 22.6 & 13.2 & 17.6 & 12.6 \\
$P_{AC}(linacs) [MW]$ & 97 & 132 & 141 & 220 \\ 
 \hline 
$G_{acc}$ [MV/m] & 23.5 & 48 & 50.2 & 36 \\
$L_{tot}$ [km] & 33 & 30 & 16 & linac 19 \\ \hline 
\end{tabular}
\end{center}
\end{table*}

\begin{figure}
\epsfysize200pt
\begin{sideways}
\figurebox{}{}{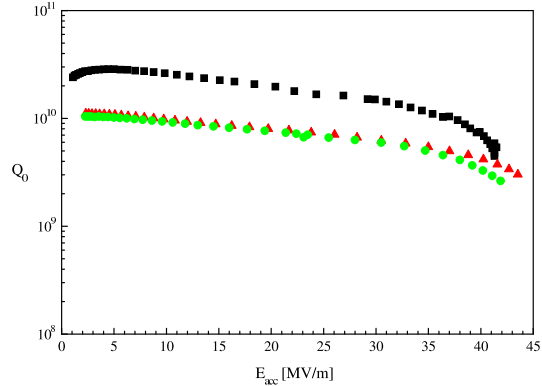}
\end{sideways}
\caption{Excitation curves of three electropolished single-cell
cavities. Gradients well above 35 MV/m are reached.}
\label{fig:elpol}
\end{figure}

Table~\ref{tab:comp} also contains the presently planned length of the
facilities~\cite{tesla,nlc,xband,cband}. An upgrade in energy up to around
one TeV seems possible for all designs. In the NLC case, more cavities would be
installed within the existing tunnel, in the JLC case, the tunnel length
would have to be increased to house more cavities. In the TESLA case,
a gradient of around 35 MV/m is needed to reach $\sqrt{s}$ of 800 GeV
within the present tunnel length. Higher energies would probably require
an extension of the tunnel.
Such gradients have repeatedly been reached
in tests of single-cell cavities whose surfaces have been electropolished not
only chemically treated. The result of this common effort from
KEK, CERN, Saclay and DESY is shown in figure~\ref{fig:elpol}. \\

In summary, all designs are very well advanced. The TESLA collaboration
has presented a fully costed Technical Design Report in March 2001. The 
other collaborations are expected to provide such reports within the
next years. The construction of a linear collider with at least 500 GeV
centre-of-mass energy, with upgrade potential to around one TeV,
could start soon.

\subsection{Multi-TeV Range}

\begin{figure*}
\epsfxsize200pt
\begin{sideways}\begin{sideways}\begin{sideways}
\figurebox{}{200pt}{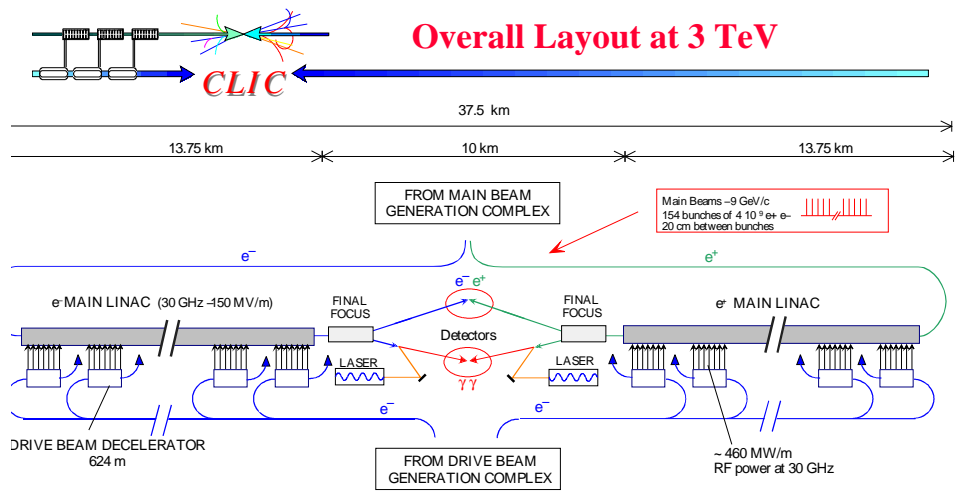}
\end{sideways}\end{sideways}\end{sideways}
\caption{Overall layout of the CLIC complex for a centre-of-mass energy
of 3 TeV.}
\label{fig:clic}
\end{figure*}

To reach centre-of-mass energies beyond the TeV range, up to 3-5 TeV, a 
two beam acceleration
concept (CLIC) with very high accelerating fields is being developed at 
CERN~\cite{wilson}. The schematic layout of that facility is shown in
figure~\ref{fig:clic}. It is optimised for $\sqrt{s}$ of 3 TeV, using high
frequency (30 GHz) normalconducting structures operating at very high
accelerating fields (150 MV/m). The present design calls for bunch 
separations of .67 ns, a vertical spotsize of 1 nm and beamstrahlung 
$\delta_E$ of 30\%. For this promising concept
a new test facility is under construction at CERN which
should allow tests with full gradient starting in 2005.  

\section{Realisation}

The new generation of high energy colliders most likely
exceeds the resources of a country or even a region. 
There is general consensus that the realisation has to be
done in an international, interregional framework. One such
framework, the so called 
Global Accelerator Network (GAN), has been proposed to ICFA in March 2000. 
A short discussion of the principle considerations will be presented
here, more details can
be found in ref.~\cite{teslagan}.

The GAN is a global collaboration of laboratories and institutes
in order to design, construct, commission, operate and
maintain a large accelerator facility. The model is based on the
experience of large experimental collaborations, particularly in
particle physics. Some key elements are listed below:
\begin{itemize}
\item it is not an international permanent institution, but an international
project of limited duration;
\item the facility would be the common property of the participating countries;
\item there are well defined roles and obligations of all partners;
\item partners contribute through components or subsystems;
\item design, construction and testing of components is done in
participating institutions;
\item maintenance and running of the accelerator would be done to a large
extent from the participating institutions.
\end{itemize}

The GAN would make best use of worldwide competence, ideas and resources,
create a visible presence of activities in all participating countries
and would, hopefully, make the site selection less controversial.

ICFA has set up two working groups to study general considerations of
implementing a GAN and to study the technical considerations and influence
on the design and cost of the accelerator. The reports of these working
groups can be found on the web~\cite{icfa}. Their overall conclusion is
that a GAN can be a feasible way to build and operate a new global
accelerator, although many details still need to be clarified.  

\section{Summary}

There is global consensus about the next accelerator based project
in particle physics. It has to be an electron-positron linear
collider with an initial energy reach of some 500 GeV with the potential
of an upgrade in centre-of-mass energy. The physics case is excellent,
only a few highlights could be presented here.
There is also global consensus that concurrent operation with LHC is
needed and fruitful. Therefore, a timely realisation is mandatory.
The technical realisation of a linear collider is now feasible,
several technologies are either ripe or will be ripe soon. A fast consensus
in the community about the technology is called for
having in mind
a timely realisation as a global project with the highest possible
luminosity and a clear upgrade potential beyond 500 GeV. 

\section*{Acknowledgments}
The author would like to express his gratitude to all people who
have contributed to the studies of future electron-positron linear
colliders from the machine design to physics and detector studies. 
Special thanks go to the organisers and their team
for a very well organised, inspiring conference as well as for the competent
technical help in preparing this presentation.


\begin{thebibliography}{99}

\bibitem{ecfaphys}

J.A. Aguilar-Saavedra et al, TESLA Technical Design Report, Part III,
{\it Physics at an $e^+e^-$ Linear Collider},
DESY 2001-011, ECFA 2001-209, hep-ph/0106315.

\bibitem{snowmass}
T. Abe et al,
{\it Linear Collider Physics Resource Book for Snowmass 2001},
BNL-52627, CLNS 01/1729, FERMILAB-Pub-01/058-E, LBNL-47813,
SLAC-R-570, UCRL-ID-143810-DR, LC-REV-2001-074-US, 
hep-ex/0106055-58

\bibitem{jlcphys}
K. Abe et al,
{\it Particle Physics Experiments at JLC},
KEK-Report 2001-11, hep-ph/0109166.

\bibitem{fnal}

Proceedings of LCWS, {\it Physics and Experiments with Future Linear 
Colliders}, eds A. Para, H.E. Fisk, (AIP Conf. Proc., Vol 578, 2001).

\bibitem{world}
{\it Worldwide Study of the Physics and Detectors for Future $e^+e^-$
Colliders}\\
http://lcwws.physics.yale.edu/lc/

\bibitem{acfa}
{\it ACFA Joint Linear Collider Physics and Detector Working Group}\\
http://acfahep.kek.jp/

\bibitem{ecfa}
{\it 2nd Joint ECFA/DESY Study on Physics and Detectors for a Linear
Electron-Positron Collider}\\ 
http://www.desy.de/conferences/ecfa-desy-lc98.html

\bibitem{usweb}
{\it A Study of the Physics and Detectors for Future
Linear $e^+e^-$ Colliders: American Activities}\\
http://lcwws.physics.yale.edu/lc/ameri\-ca.html

\bibitem{det}

G. Alexander et al, TESLA Technical Design Report, Part IV,
{\it A Detector for TESLA},
DESY 2001-011, ECFA 2001-209.

\bibitem{g2}
H.~N.~Brown {\it et al.}  [Muon g-2 Collaboration],
Phys.\ Rev.\ Lett.\  {\bf 86} (2001) 2227

\bibitem{drees}
J. Drees, {\it these proceedings}

\bibitem{ATF}

E.Hinode et al, eds., KEK Internal 95-4, 1995,
eds J.Urakawa and M.Yoshioka, {\it Proceedings of the SLAC/KEK Linear
Collider Workshop on Damping Ring}, KEK 92-6, 1992

\bibitem{FFTB}

{\it The FFTB Collaboration:} \\
BINP (Novosibirsk/Protvino), DESY, FNAL,
KEK, LAL(Orsay), MPI Munich, Rochester, and SLAC

\bibitem{TTF}

{\it Proposal for a TESLA Test Facility}, DESY TESLA-93-01, 1992

\bibitem{jlc}
KEK-Report 97-1, 1997.

\bibitem{nlc}
{\it Zeroth Order Design Report for the Next Linear Collider}, SLAC
Report 474, 1996.
{\it 2001 Report on the Next Linear Collider},
Fermilab-Conf-01-075-E, LBNL-47935, SLAC-R-571, UCRL-ID-144077

\bibitem{tesla}

J. Andruszkow et al, TESLA Technical Design Report, Part II, 
{\it The Accelerator},
DESY 2001-011, ECFA 2001-209

\bibitem{napoly}

O.Napoly, {\it TESLA Linear Collider: Status Report}, in ref~\cite{fnal}

\bibitem{raub}

T.O. Raubenheimer, {\it Progress in the Next Linear Collider Design},
in ref~\cite{fnal}

\bibitem{chin}

Y.H. Chin et al {\it Status of JLC Accelerator Development}, in ref \cite{fnal}

\bibitem{snowacc}

A. Chao et al, {\it 2001 Snowmass Accelerator R\&D Report},
http://www.hep.anl.  \\
gov/pvs/dpb/Snowmass.pdf

\bibitem{xband}

Y.H. Chin, {\it private communication}

\bibitem{cband}

H.Matsumoto, T.Shintake, {\it private communication}

\bibitem{wilson}

I.Wilson, {\it A Multi-TeV Compact $e^+e^-$ Linear Collider},
in ref~\cite{fnal}

\bibitem{teslagan}

F. Richard et al, TESLA Technical Design Report, Part I, {\it Executive
Summary},
DESY 2001-011, ECFA 2001-209, hep-ph/0106314.

\bibitem{icfa}

http://www.fnal.gov/directorate/icfa/ \\
icfa\_tforce\_reports.html

\end{thebibliography}
\end{document}